\documentclass[journal]{IEEEtran}
\ifCLASSINFOpdf
   \usepackage[pdftex]{graphicx}
\else
\fi
\ifCLASSOPTIONcompsoc
  \usepackage[caption=false,font=normalsize,labelfont=sf,textfont=sf]{subfig}
\else
\usepackage[caption=false,font=footnotesize]{subfig}
\fi

\usepackage{amssymb}
\usepackage{balance}
\begin{document}
%
\title{Lossy Compression of Cellular Network KPIs}
%
%
%

\author{Andrea Pimpinella, Fabio Palmese, Alessandro E. C. Redondi
\thanks{Andrea Pimpinella (andrea.pimpinella@unibg.it) is with the Department of Management, Information and Production Engineering (DIGIP), University of Bergamo, Bergamo, Italy. F. Palmese (fabio.palmese@polimi.it) and A. E. C. Redondi ( alessandroenrico.redondi@polimi.it), are with the Department of Electronics, Information and Bioengineering, Politecnico di Milano, Milan, Italy.}
}

%
%

\markboth{Journal of \LaTeX\ Class Files,~Vol.~14, No.~8, August~2015}%
{Shell \MakeLowercase{\textit{et al.}}: Bare Demo of IEEEtran.cls for IEEE Journals}
%



\maketitle

\begin{abstract}
Network key performance indicators (KPIs) are a fundamental component of mobile cellular network monitoring and optimization. Their massive volume, resulting from fine-grained measurements collected across many cells over long time horizons, poses significant challenges for storage, transport, and large-scale analysis. In this letter, we show that common cellular KPIs can be efficiently compressed using standard lossy compression schemes based on prediction, quantization, and entropy coding, achieving substantial reductions in reporting overhead. Focusing on traffic volume KPIs, we first characterize their intrinsic compressibility through a rate–distortion analysis, showing that signal-to-noise ratios around 30 dB can be achieved using only 3–4 bits per sample, corresponding to an 8–10× reduction with respect to 32-bit floating-point representations. We then assess the impact of KPI compression on representative downstream analytics tasks. Our results show that aggregation across cells mitigates quantization errors and that prediction accuracy is unaffected beyond a moderate reporting rate. These findings indicate that KPI compression is feasible and transparent to network-level analytics in cellular systems.
\end{abstract}

\begin{IEEEkeywords}
Network KPIs, Lossy compression, Network analytics
\end{IEEEkeywords}

%
\IEEEpeerreviewmaketitle


\section{Introduction}\label{sec:intro}
Mobile cellular networks are continuously monitored through a large set of Key Performance Indicators (KPIs), collected from the Radio Access Network (RAN) and the core network. These KPIs are typically represented as floating-point time series and constitute the primary input to operational analytics such as forecasting, anomaly detection, and capacity planning. In contemporary 4G and 5G networks, KPIs are collected across thousands of network elements and cells, at fine time granularity and over long time horizons. As a result, the volume of KPI data grows rapidly, posing significant challenges in terms of storage, transport, and large-scale analysis.

Despite their massive volume, network KPI time series are typically handled using lossless compression or highly conservative lossy schemes.
This practice is largely motivated by the operational criticality of KPIs and the lack of widely accepted guidelines that relate lossy compression to the reliability of downstream network analytics.
As a result, high pointwise reconstruction fidelity is often treated as a necessary requirement, even though KPIs are rarely consumed as raw sample-level signals and are instead analyzed through aggregated statistics or predictive models.

In parallel, lossy compression of floating-point time series has been extensively studied in the literature, albeit in markedly different application domains. Prediction–quantization–entropy coding frameworks represent a dominant paradigm, as exemplified by LFZip \cite{chandak2020lfzip}, which achieves state-of-the-art rate–distortion performance under explicit error bounds on raw high-frequency measurements. More recent work has focused on compression for time-series databases, where data are processed in short slices and decompression speed is critical; Machete \cite{shi2024machete}, for instance, demonstrates substantial gains in compression ratio and query performance when integrated into InfluxDB, a popular time series database. While effective within their respective scopes, these approaches primarily target generic sensor, scientific, or system-monitoring data.

This domain gap is further confirmed by a recent comprehensive survey on time series compression \cite{chiarot2023time}, which systematizes the literature across databases, IoT, and signal processing. The survey highlights that existing benchmarks and evaluation methodologies predominantly rely on generic datasets and focus on reconstruction fidelity, compression ratio, or system-level efficiency. Notably, network performance indicators and mobile cellular workloads are absent, and to the best of our knowledge, the impact of compression on downstream analytical tasks is not explicitly considered.

This limitation is particularly relevant for network KPIs, which are inherently aggregated across space and time and are primarily used as inputs to downstream analytics operating on temporal trends, such as traffic forecasting or capacity planning. However, the relationship between compression rate, aggregation, and downstream task performance has not been systematically investigated for network KPI time series.
Existing approaches, such as the Quality of Monitoring (QoM)
framework \cite{motlagh2021quality}, focus on controlling KPI-level information loss and reconstruction accuracy, but do not explicitly assess how compression impacts downstream analytics operating on aggregated KPIs or forecasting tasks.

In this paper, we adopt a task-centric perspective and study lossy compression of network KPIs through the lens of rate--distortion analysis. 
Rather than focusing solely on pointwise reconstruction error, we explicitly characterize how bitrate reduction affects representative downstream analytics, including spatial aggregation of KPIs across multiple cells and short-term traffic forecasting. Our study considers three widely used cellular KPIs—downlink traffic volume, physical resource block (PRB) occupancy, and number of active users—which jointly capture traffic demand, resource utilization, and user activity. By combining rate–distortion analysis with task-level evaluation, we provide a unified view of compression efficiency and analytical performance for network KPI time series.

\section{Methodology}
\subsection{Dataset}
Our analysis is based on operational measurements collected from approximately 3000 LTE (4G) cells deployed in a mid-size European city. The dataset spans one month (October 2023) and consists of hourly sampled time series. For each cell, we consider three widely used radio access network KPIs: downlink traffic volume, physical resource block (PRB) occupancy, and number of active users. These KPIs capture complementary aspects of network behavior related to traffic demand, resource utilization, and user activity, and are routinely monitored in operational cellular networks.

\subsection{KPI Time-Series Model}
Let $\mathcal{C}$ denote the set of cells and consider a given KPI.
For each cell $c \in \mathcal{C}$, we denote by $x_c[n]$ the corresponding discrete-time KPI time series, sampled at uniform hourly intervals.
Time series containing missing samples (NaNs) are discarded before the analysis.

\subsection{Quantization Model}

All compression schemes considered in this work rely on uniform scalar quantization.
We denote by $Q_\Delta(\cdot)$ a uniform scalar quantizer with step size $\Delta$, defined as
\begin{equation}
Q_\Delta(u) = \Delta \cdot \mathrm{round}\!\left(\frac{u}{\Delta}\right).
\end{equation}
The quantization step size $\Delta$ is varied to trace the rate--distortion characteristics of each representation.

\subsection{Prediction and Transform-Domain Representations}

To characterize the intrinsic rate--distortion behavior of network KPIs, we consider a set of classical sample-domain and transform-domain compression schemes, applied independently to each cell time series.

\begin{itemize}

\item{\textit{Pulse-code modulation (PCM)}}: PCM serves as a baseline and applies no prediction or transform.
Quantization is performed directly in the sample domain:
\begin{equation}
\hat{x}_c[n] = Q_\Delta\!\left(x_c[n]\right).
\end{equation}

\item{\textit{Differential PCM (DPCM)}}: DCPM exploits temporal correlation through first-order prediction.
The prediction residual is quantized and reconstruction is performed in closed loop:
\begin{equation}
\hat{x}_c[n] = \hat{x}_c[n-1] + Q_\Delta\!\left(x_c[n] - \hat{x}_c[n-1]\right),
\end{equation}
with $\hat{x}_c[0]$ initialized to the first sample.

\item{\textit{Block discrete cosine transform (DCT)}}: DCT is applied to weekly KPI vectors.
Specifically, non-overlapping blocks of $168$ samples (corresponding to one week of hourly measurements) are transformed.
The DCT is a fixed, closed-form orthonormal transform that does not depend on the data distribution.
Let $\mathbf{D} \in \mathbb{R}^{168 \times 168}$ denote the DCT matrix.
Transform coefficients are quantized independently and reconstruction is obtained via the inverse transform:
\begin{equation}
\hat{\mathbf{x}}_c = \mathbf{D}^\top Q_\Delta\!\left(\mathbf{D}\mathbf{x}_c\right).
\end{equation}
 \item{\textit{Karhunen--Loève transform (KLT)}}: we finally consider the KLT, which optimally decorrelates the KPI samples [X].
Unlike the DCT, the KLT is data-driven and must be estimated from a set of training time series.
Specifically, the transform is derived from the empirical covariance matrix
\begin{equation}
\mathbf{R} = \mathbf{X}_{\mathrm{tr}}\mathbf{X}_{\mathrm{tr}}^\top,
\end{equation}
where $\mathbf{X}_{\mathrm{tr}}$ collects weekly KPI vectors from a subset of cells randomly selected from $\mathcal{C}$.
Let $\mathbf{R} = \mathbf{V}\mathbf{\Lambda}\mathbf{V}^\top$ denote its eigendecomposition.
Compression is performed by quantizing the decorrelated coefficients:
\begin{equation}
\hat{\mathbf{x}}_c = \mathbf{V} Q_\Delta\!\left(\mathbf{V}^\top \mathbf{x}_c\right).
\end{equation}
\end{itemize}

\subsection{Rate--Distortion Metrics}

Reconstruction error is measured over all cells and time samples using the mean squared error (MSE) and reported in terms of signal-to-noise ratio (SNR), defined as
\begin{equation}
\mathrm{SNR} = 10 \log_{10}\left(\frac{\mathrm{Var}(x_c(n))}{
\mathbb{E}\!\left[(x_c(n)-\hat{x}_c(n))^2\right]}\right),
\end{equation}
where variance and expectation are computed over all cells $c \in \mathcal{C}$ and time indices $n$.

Rather than implementing a specific entropy coder, we assume \emph{ideal entropy coding} and estimate the achievable rate from the empirical distribution of the quantization indices.
Let $p_i$ denote the empirical probability mass function of the indices produced by a given scheme.
The corresponding coding rate is estimated as
\begin{equation}
R = -\sum_i p_i \log_2 p_i,
\end{equation}
yielding a rate expressed in bits per sample.

By sweeping $\Delta$, this procedure provides the rate--distortion characteristics of each representation, independently of implementation-specific coding choices.

\begin{figure*}[t]
    \centering
    \subfloat[PRB occupancy]{
        \includegraphics[width=0.3\textwidth]{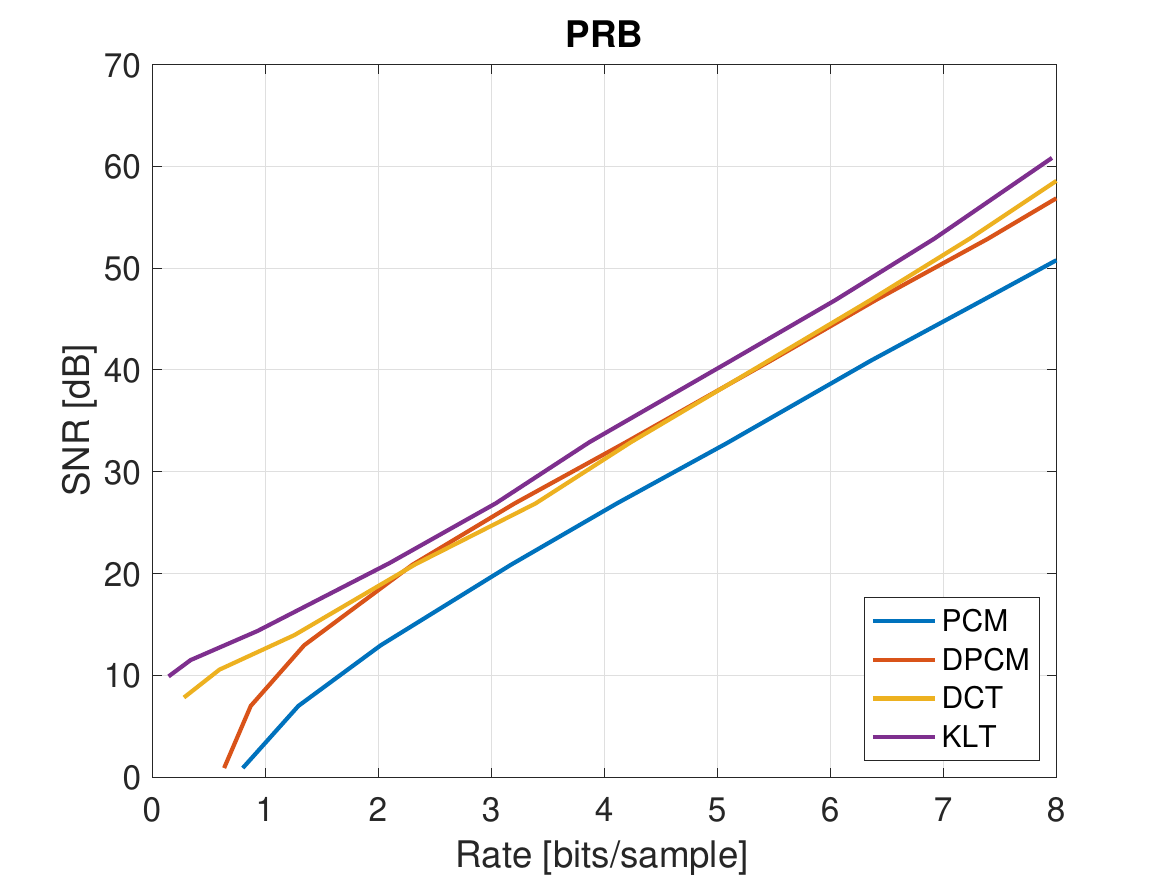}
        \label{fig:rd_prb}
    }
    \hfill
    \subfloat[Active users (RRC)]{
        \includegraphics[width=0.3\textwidth]{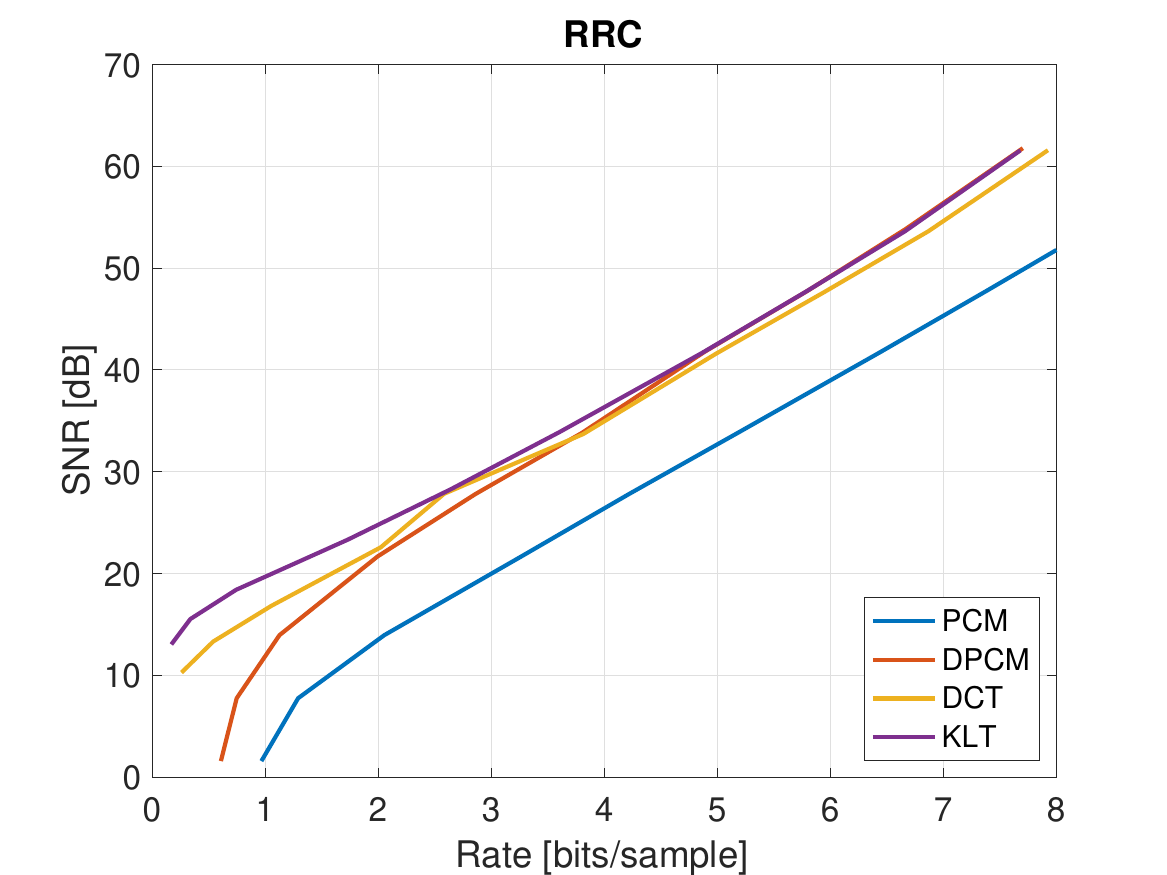}
        \label{fig:rd_rrc}
    }
    \hfill
    \subfloat[Downlink traffic volume]{
        \includegraphics[width=0.3\textwidth]{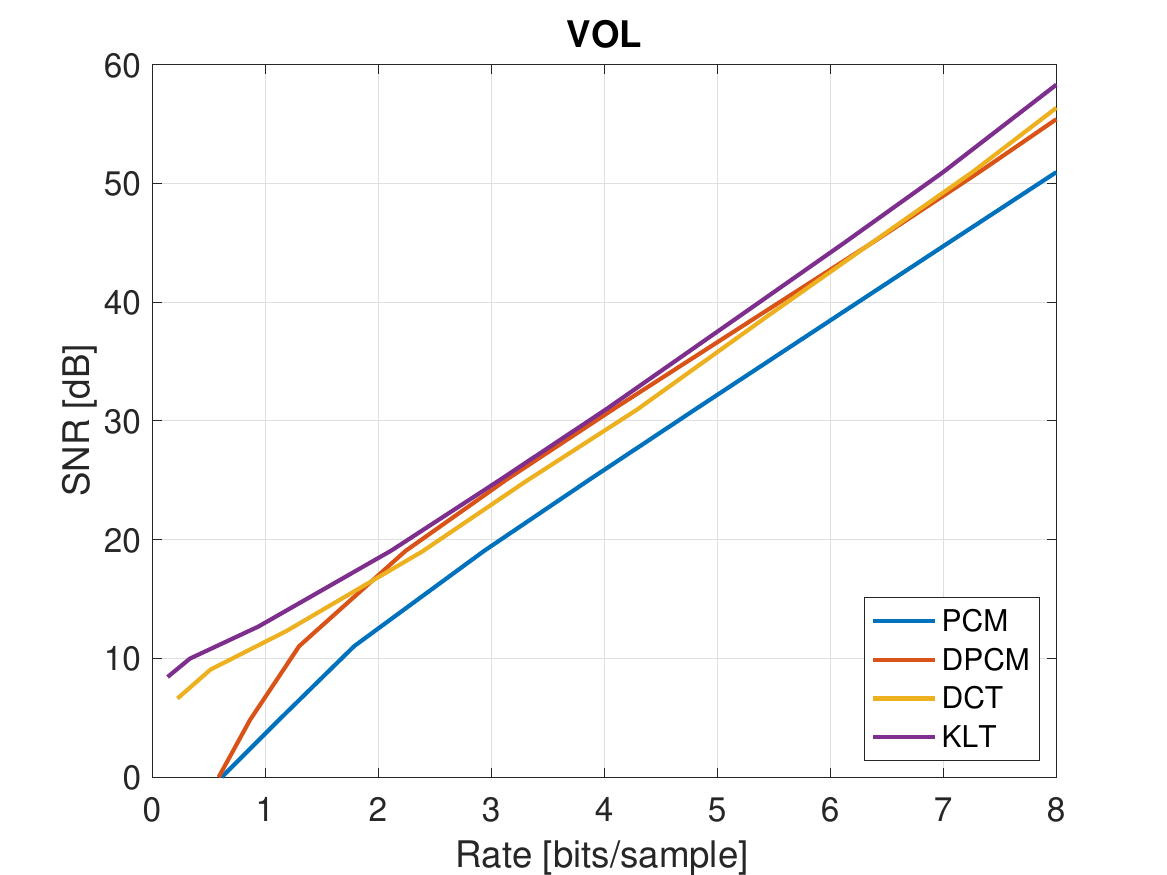}
        \label{fig:rd_vol}
    }
    \caption{Rate--distortion characteristics for the three considered KPIs. SNR is reported as a function of the entropy-estimated rate for PCM, DPCM, DCT, and KLT.}
    \label{fig:rd_all}
\end{figure*}

\section{Rate--Distortion Analysis}

Fig.~\ref{fig:rd_all} reports the rate--distortion characteristics of the considered compression schemes for the three KPIs under study: PRB occupancy, number of active users (RRC), and downlink traffic volume.
Across all cases, transform-based representations substantially outperform sample-domain schemes, confirming the strong temporal correlation present in network KPI time series.

Direct sample quantization (PCM) exhibits the poorest rate--distortion performance across all three KPIs.
DPCM consistently improves upon PCM by exploiting short-term temporal correlation and achieves reasonable performance over a broad range of bitrates.
Nevertheless, transform-based schemes provide superior rate--distortion efficiency overall.

Among the considered representations, the KLT consistently achieves the highest SNR at a given bitrate, as expected from its data-driven construction.
The performance advantage of KLT over DPCM is present across the entire rate range and becomes particularly pronounced in the low-bitrate regime.
DCT closely follows the KLT across all three KPIs, capturing most of the available compression gain without requiring any training phase.

As a result, while the KLT offers the best rate--distortion performance, its modest advantage over the closed-form DCT may not always justify the additional complexity associated with estimating the transform in practical deployments.

From a practical point of view, the most relevant operating region lies between approximately $2$ and $4$~bit/sample.
In this range, transform-based schemes already achieve high reconstruction quality (on the order of $20$--$30$~dB in SNR), while corresponding to a compression factor of roughly $8\times$ to $16\times$ with respect to standard 32-bit floating-point representations.
In contrast, PCM remains significantly less efficient at comparable bitrates.
At moderate to high bitrates, the rate--distortion curves exhibit an approximately linear behavior, consistent with the high-rate quantization regime.

\section{Impact on Downstream Analytics}
\label{sec:downstream}

In this section, we evaluate the impact of transform-based compression on representative downstream analytics tasks commonly employed in cellular network monitoring and optimization.

\subsection{Aggregated KPI Accuracy}
\label{subsec:aggregation}

We first assess the impact of compression on aggregated KPIs that are
directly relevant at the core network level.
Let $\mathbf{x}_c \in \mathbb{R}^T$ and $\hat{\mathbf{x}}_c$ denote the
original and reconstructed volume KPI time series for cell
$c \in \mathcal{C}$, respectively.
In our case study, aggregation is performed by summation across cells,
which corresponds to the total traffic observed at the core network and
represents a key quantity for network monitoring and capacity planning. Formally, the aggregated traffic obtained from the original KPIs is given by $\mathbf{s} = \sum_{c \in \mathcal{C}} \mathbf{x}_c$, while aggregation of the reconstructed KPIs yields $\hat{\mathbf{s}} = \sum_{c \in \mathcal{C}} \hat{\mathbf{x}}_c$. 

\begin{figure*}[t]
    \centering
    \subfloat[PRB occupancy]{
        \includegraphics[width=0.3\textwidth]{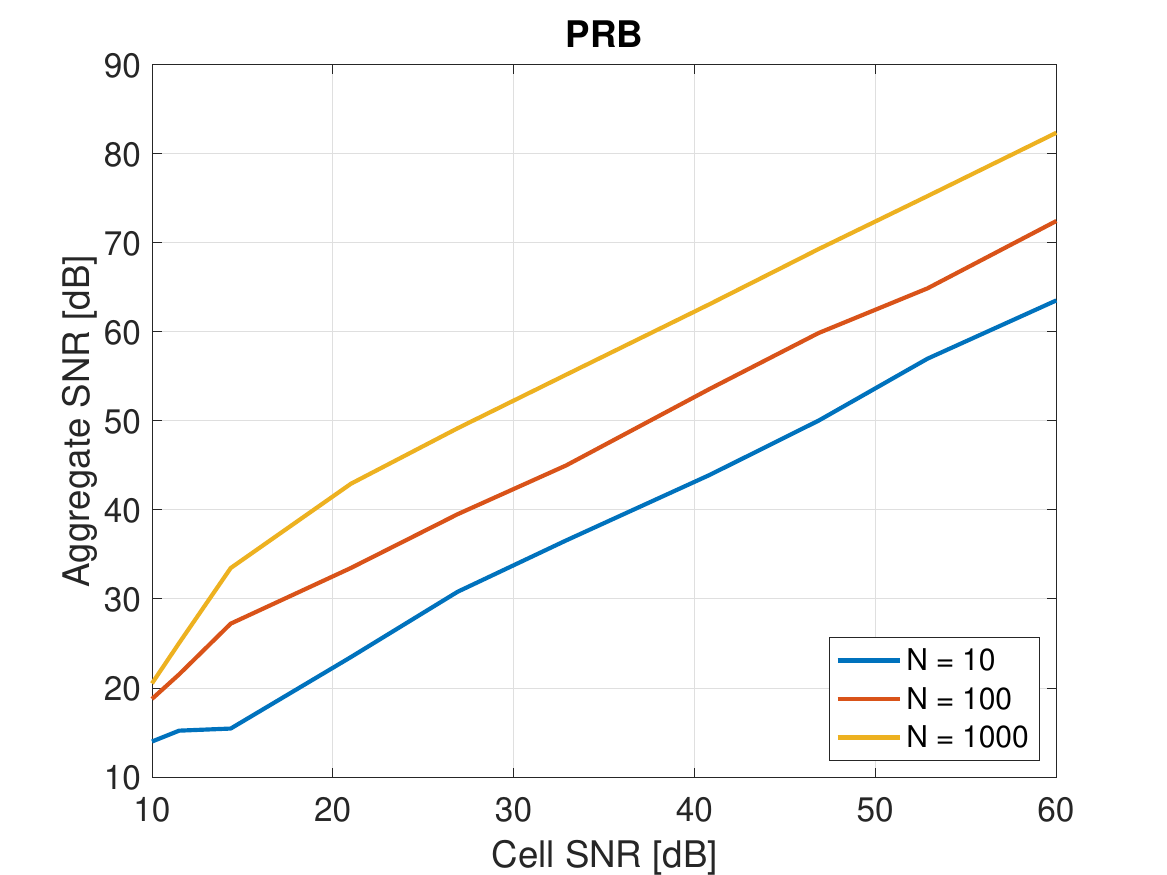}
        \label{fig:agg_prb}
    }
    \hfill
    \subfloat[Active users (RRC)]{
        \includegraphics[width=0.3\textwidth]{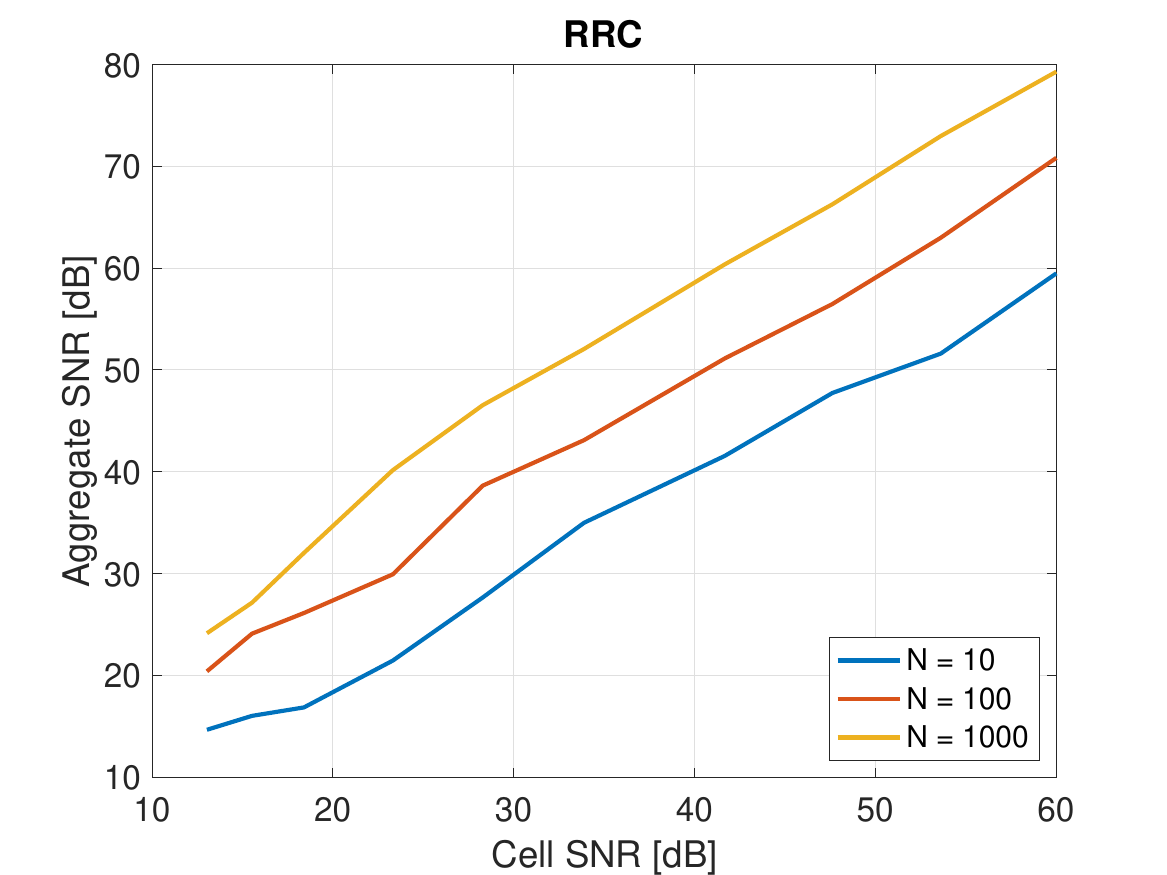}
        \label{fig:agg_rrc}
    }
    \hfill
    \subfloat[Downlink traffic volume]{
        \includegraphics[width=0.3\textwidth]{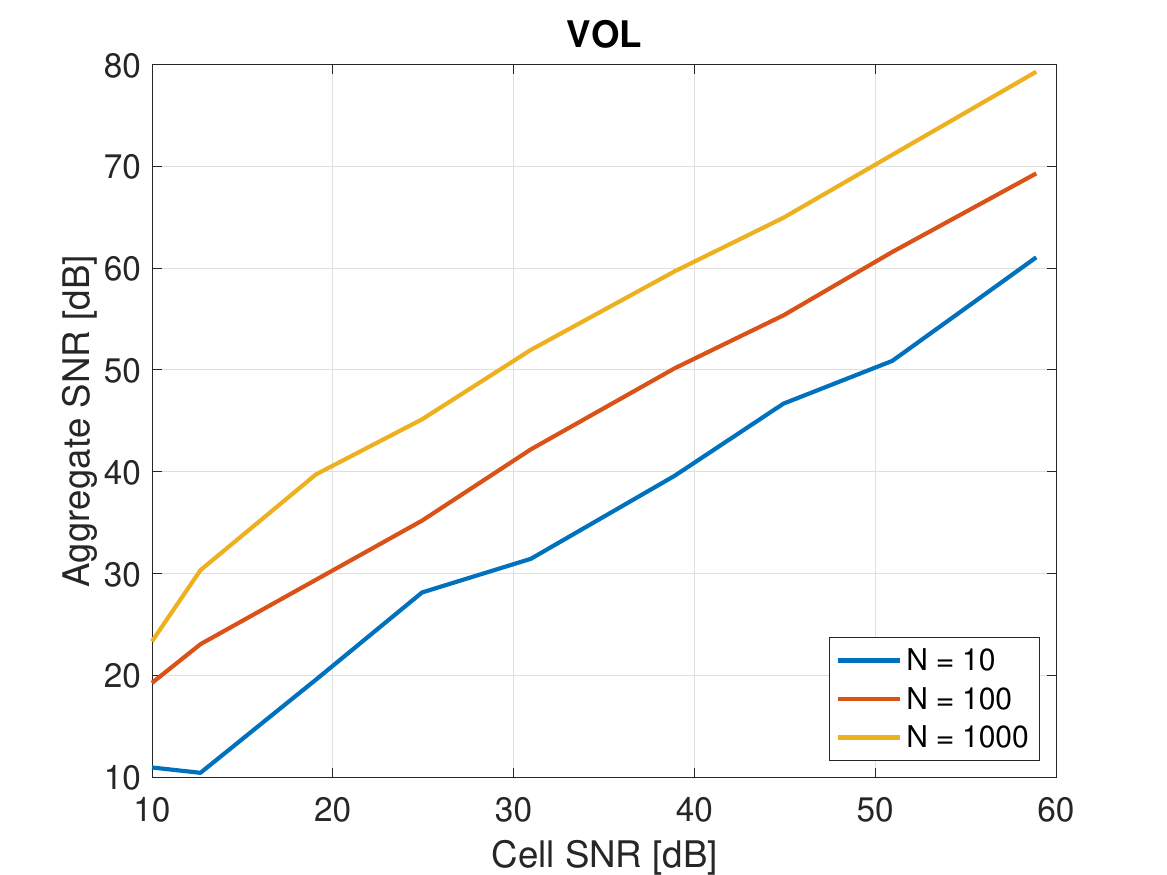}
        \label{fig:agg_vol}
    }
    \caption{Aggregate SNR as a function of the per-cell SNR for three
representative KPIs: (a) PRB, (b) Active Users (RRC), and (c) Volume (VOL).
Results are shown for different numbers of aggregated cells
($N=10,100,1000$).
In all cases, aggregation by summation yields a substantial SNR gain that
increases with $N$, reflecting the attenuation of uncorrelated
quantization noise across cells.}
    \label{fig:agg}
\end{figure*}

We quantify the effect of compression by measuring the signal-to-noise
ratio (SNR) between $\mathbf{s}$ and $\hat{\mathbf{s}}$.
Fig.~\ref{fig:agg} reports the aggregate SNR as a function of the
average per-cell SNR for different numbers of aggregated cells
($N = 10, 100, 1000$).
The results show that aggregation yields a substantial SNR gain with
respect to the cell-level reconstruction quality.
In particular, when aggregating $N=1000$ cells, an aggregate SNR of
approximately $30$~dB is achieved even when the average per-cell SNR is
only about $15$~dB.
This corresponds to operating points where transform-based compression
schemes such as KLT or DCT require only $1$--$2$~bit/sample to achieve the
necessary cell-level reconstruction quality.
As $N$ increases, the aggregate SNR improves further, confirming that
summation across a large number of cells strongly attenuates
quantization noise while preserving the dominant traffic components.

This behavior highlights that summation across cells strongly attenuates
quantization noise: reconstruction errors add incoherently across cells,
whereas the useful traffic components combine coherently.
As a result, compression operating points that appear moderate at the
cell level translate into high-fidelity representations of aggregated
traffic at the core network.

\begin{figure*}[t]
    \centering
    \subfloat[PRB occupancy]{
        \includegraphics[width=0.3\textwidth]{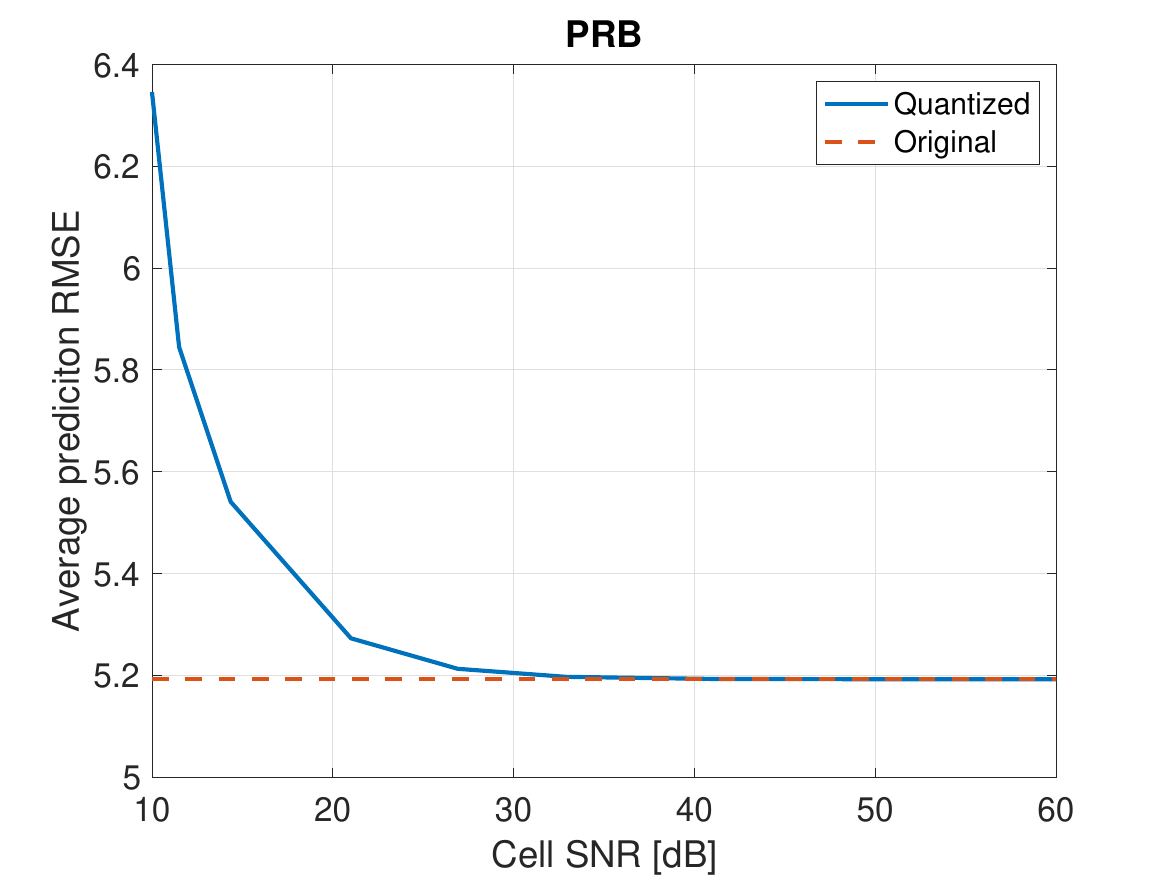}
        \label{fig:mws_prb}
    }
    \hfill
    \subfloat[Active users (RRC)]{
        \includegraphics[width=0.3\textwidth]{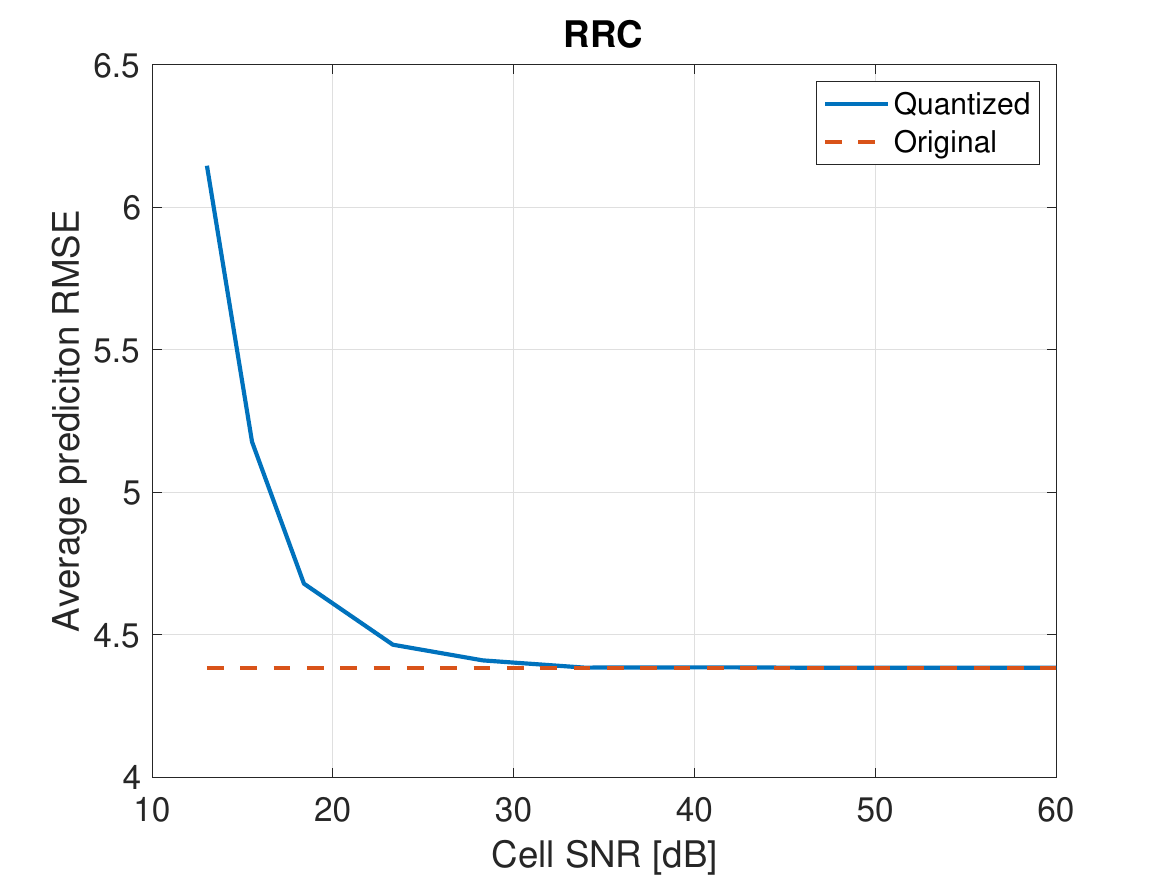}
        \label{fig:mws_rrc}
    }
    \hfill
    \subfloat[Downlink traffic volume]{
        \includegraphics[width=0.3\textwidth]{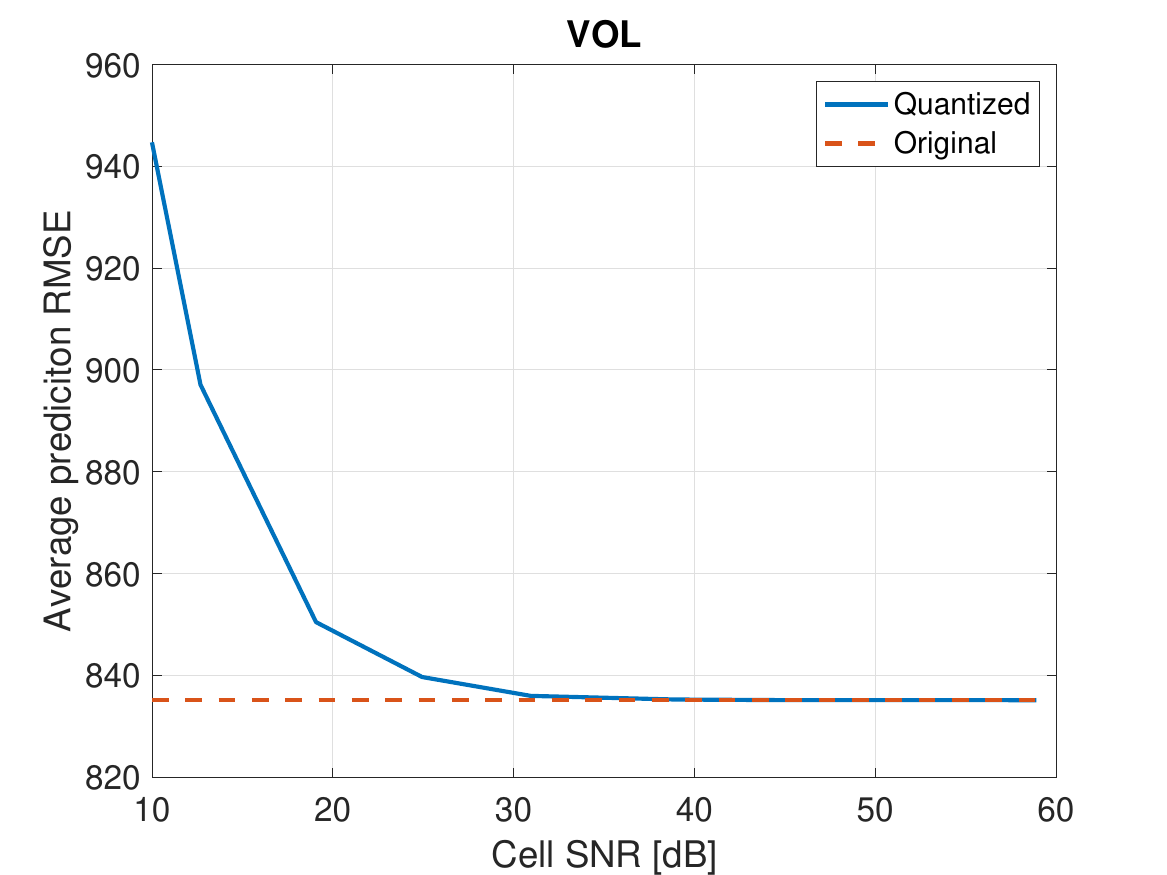}
        \label{fig:mws_vol}
    }
\caption{Forecasting performance using the Median Weekly Signature (MWS)
predictor.
Average RMSE as a function of the per-cell SNR for three representative
KPIs: (a) PRB occupancy, (b) number of active users (RRC), and (c)
downlink traffic volume.
Errors are reported in KPI-native units (percentage for PRB, number of
users for RRC, and MB/hour for volume).
Results are shown for original and KLT-quantized data.
Across all KPIs, forecasting accuracy remains stable once a moderate
cell-level SNR is achieved.}
    \label{fig:forecasting}
\end{figure*}

\subsection{Forecasting Performance}
\label{subsec:forecasting}

We next evaluate the impact of compression on a downstream forecasting
task for multiple network KPIs.
For this experiment, compressed KPIs are obtained exclusively using
KLT-based transform coding, which provides the best rate--distortion
performance among the considered schemes.
This choice allows us to focus on the effect of quantization on
forecasting accuracy under the most favorable compression setting.
Specifically, we consider the prediction of future KPI values using the
Median Weekly Signature (MWS), which is widely used in mobile traffic
analysis and forecasting as it captures stable weekly usage patterns
while being robust to outliers \cite{furno2016tale}.
Given three consecutive weeks of historical KPI data, the weekly
signature is computed as the sample-wise median across the three weeks
and is then used to forecast the corresponding samples of the fourth
week.
The prediction is performed independently for each cell and for each
KPI, i.e., one MWS is estimated per cell and per KPI.
Forecasting performance is evaluated by computing the root mean square
error (RMSE) for each cell and KPI and averaging the results across all
cells.
Forecasts are generated starting from either the original KPI time series
$\mathbf{x}_c$ or their compressed and reconstructed counterparts
$\hat{\mathbf{x}}_c$.
Prediction accuracy is evaluated as a function of the per-cell
reconstruction quality, measured in terms of SNR.
Fig.~\ref{fig:forecasting} reports the forecasting RMSE as a function of
the cell-level SNR for three representative KPIs (VOL, PRB, and RRC),
comparing original and KLT-quantized data.
Fig.~\ref{fig:forecasting} shows that, for per-cell SNR values above
approximately $30$~dB, the forecasting RMSE obtained from KLT-quantized KPIs closely matches that obtained from the original data for all
considered KPIs.
In this regime, no measurable degradation in forecasting accuracy is observed as a result of compression.

\section{Conclusions}
\label{sec:conclusions}

This paper investigated lossy compression of cellular network KPIs from
a rate--distortion and task-centric perspective.
Our results show that lossy compression is not only feasible but also
highly effective for network analytics.
In particular, operating points corresponding to approximately
$30$~dB of per-cell reconstruction quality can be achieved at
$3$--$4$~bit/sample using transform-based schemes, while aggregation at
the core network enables reliable representations at even lower rates,
on the order of $1$--$2$~bit/sample.

We further demonstrated that such compression levels have a negligible
impact on representative downstream tasks, including aggregation of
traffic at the core network and forecasting using the MWS predictor.
These findings indicate that moderate cell-level distortion translates
into high-fidelity representations of aggregated KPIs and preserves the
temporal structures exploited by downstream analytics.

Future work will explore more advanced compression strategies that
explicitly leverage spatial redundancy across cells as well as
learning-based approaches tailored to network KPIs.

\balance
\section*{Acknowledgment}
This study was carried out within the PRIN project COMPACT and received funding from Next Generation EU, Mission 4 Component 1, CUP: D53D23001340006.


\ifCLASSOPTIONcaptionsoff
  \newpage
\fi



\bibliographystyle{IEEEtran}
\bibliography{IEEEabrv,bibliography}
\end{document}